\documentstyle[aps,preprint,epsfig]{revtex}
\input {epsf}

\evensidemargin=\oddsidemargin
\begin{document}
\begin{flushright}
TUIMP-TH-97/81\\
\end{flushright}
\vspace{0.2cm}
\begin{center}
{\Large \bf Testing Technicolor Models in Top Quark Pair Production at
High Energy Photon Colliders}

\null
\vspace{0.2cm}

{\bf Hong-Yi Zhou$^{a,b}$~~Yu-Ping Kuang$^{a,b}$~~Chong-Xing Yue$^{a,c}$,\\
Hua Wang$^{a,b}$,~~ Gong-Ru Lu $^{a,c}$} ,\\
\vspace{0.6cm}  
{\small\it a. CCAST (World Laboratory), P. O. Box 8730, Beijing 100080, China}\\
{\small\it b. Institute of Modern Physics,Tsinghua University, Beijing, 
 100084, P. R. of China.\footnote{Mailing address} } \\
{\small\it c.  Physics Department, Henan Normal University, Xin Xiang, 
 Henan 453002,  P. R. of China }\\
\end{center}
\null
\vspace{-0.4cm}

\begin{abstract}
  We study pseudo-Goldstone boson corrections to $~\gamma\gamma\to
t\bar{t}~$ production rates in technicolor models with and without
topcolor at the $\sqrt{s}=0.5~{\rm and}~1.5$~TeV photon colliders. 
We find that, for reasonable ranges of the parameters, the corrections  are 
large enough to be observable, and the corrections in models with topcolor 
are considerably larger than those in models without topcolor, and they are 
all significantly larger than the corresponding corrections in the minimal 
supersymmetric standard model (MSSM) with $\tan\beta\geq 1$. So that the two 
kinds of technicolor models and the MSSM with $\tan\beta\geq 1$ can be 
experimentally distinguished.
\end{abstract}
\null
\vspace{0.1cm}

\newpage
\baselineskip=.38in
\begin{center}
{\bf I. Introduction}
\end{center}

  The electroweak symmetry breaking (EWSB) mechanism remains an open question 
in spite of the success of the standard model (SM) compared with the new 
LEP precision measurement data. In the SM, elementary Higgs field is assumed
to be in charge of the EWSB. So far the Higgs boson is not found, and theories
with elementary scalar fields suffers from the problems of triviality, 
unnaturalness, etc. Therefore studying EWSB mechanisms other than the
simple SM Higgs sector is one of the interesting topics in current particle
theory. Technicolor (TC) theory \cite{TC} is an attractive idea of dynamical 
EWSB which avoids the shortcomings arising from elementary scalar fields,
and it has been enlarged to the extended technicolor (ETC) theory \cite{ETC}
for giving masses not only to the weak gauge bosons but also to quarks and 
leptons. A series of improved ideas, such as walking technicolor (WTC) theory
\cite{WTC}\cite{AT}, multiscale walking technicolor theory \cite{MWTC}, and 
topcolor-assisted technicolor (TOPCTC) theory \cite{TOPC}, have been proposed 
to overcome the phenomenological difficulties in the ETC theory, and these 
make the theory one of the important candidates of promising mechanisms for 
EWSB. It is thus interesting to study the effects of this kind of theory in 
various physical processes and see if they can be experimentally tested. It is 
important to notice that in most of the currently interesting improved TC 
models, the TC sectors are non-minimal, i.e. they all contain certain pseudo 
Goldstone bosons (PGB's) in the few hundred GeV region. This can be seen as 
the characteristic features of this kind of models. Thus studying the effects 
of the PGB's in processes at high energy colliders will be of special interest.

  The recently discovered top quark is the heaviest particle yet
experimentally found. Its mass, $~m_t=176~$GeV \cite{CDFD0}, is of the order 
of the EWSB scale $~v = (\sqrt{2} G_F)^{-1/2} = 246~$GeV. This means that the 
top quark couples rather strongly to the EWSB sector so that the effects from 
new physics would be more apparent in processes with the top quark than with 
any other light quarks. Experimentally, it is possible to separately 
measure various production and decay form factors of the top quark 
at the level of a few percent \cite{Peskin}. Thus theoretical calculations of 
various corrections to the production and decay of the top quarks are of 
much interest. 

Top quark pair can be produced at various high energy colliders.
Of special interest is to examine the ability of the suggested future
TeV energy photon colliders in probing the EWSB mechanism via $t\bar{t}$ 
production. This paper is devoted to this kind of study. 
There have been various studies of probing the EWSB mechanism
via top quark pair productions at high energy colliders. For example, model-
independent studies \cite{Yuan}, studies of the top quark pair
production cross sections in photon collisions in the SM, the two-Higgs-
doublet model, and the minimal supersymmetric SM (MSSM) \cite{ttbar}\cite{DDS}
\cite{MSSM}, and the study of PGB contributions to the $t\bar{t}$ production 
cross sections at the Fermilab Tevatron \cite{EL} and the CERN LHC in a 
topcolor-assisted multiscale technicolor model (TOPCMTC) \cite{Lane}
\cite{YZKL}, etc. The results in Ref.\cite{YZKL} show that PGB's do give 
significant contributions to the $t\bar{t}$ production cross section from 
which the color-octet neutral technipion and the neutral top-pion with mass 
close to $350$~GeV can be clearly tested at the LHC and the Tevatron for 
reasonable range of the parameters in the TOPCMTC model. In this paper we 
study the PGB contributions to the $~\gamma\gamma\to t\bar{t}~$ cross section 
at the $\sqrt{s}=0.5~{\rm and}~1.5$~TeV photon colliders in various 
technicolor models. We shall show that, for reasonable values of the 
paremeters in the models, the PGB contributions are quite large in models 
assisted by topcolor, and are considerably smaller in models without topcolor. 
All these cross sections are significantly larger than those in the MSSM for 
$\tan\beta\geq 1$ \cite{MSSM}. So that different models can be distinguished 
by the $\gamma\gamma\to t\bar{t}$ cross section measurement at the high energy 
photon colliders. At the $\sqrt{s}=1.5$~TeV photon collider, even the original 
TOPCTC model and the TOPCMTC model can be experimentally ditinguished.

 This paper is organized as follows. In Sec.II, we take the Appelquist-Terning 
one-family WTC model \cite{AT} as a typical example of the reasonable TC 
models without assisted by topcolor, and present the results of the PGB 
contributions to the $\gamma\gamma\to t\bar{t}$ cross section at the 
$\sqrt{s}=0.5~{\rm and}~1.5$~TeV photon colliders in this model. Sec.III  
contains the corresponding results of two typical TOPCTC models, namely the 
original TOPCTC model by Hill \cite{TOPC} and the TOPCMTC model \cite{Lane}
\cite{YZKL}. Discussions and conclusions are given in Sec.IV, and the analytic 
formulae for the form factors in the production amplitudes in terms of the 
well known standard notions of one-loop Feynman integrals\cite{PV} 
are presented in the APPENDIX.

\null
\vspace{0.5cm}
\begin{center}
{\bf II. $~t\bar{t}$ production cross section in the one family WTC model} 
\end{center}

 In this section, we take the Appelquist-Terning one-familty WTC model 
\cite{AT} as a typical example of reasonable TC models without assisted by 
topcolor to calculate the $\gamma\gamma\to t\bar{t}$ cross section. In
this model, the technilepton sector does not respect the custodial symmetry 
$~SU(2)_c~$ which makes the the oblique correction parameter $S$ \cite{PT} 
small as required by the experiment. The TC group in this model is taken to be 
$~SU(2)_{TC}~$ which minimizes the $~S~$ paremeter. There are 36 TC PGB's
composed of weak $~SU(2)_W~$ doublets of techniquarks Q and technileptons L. 
The relevant PGB's in this study are the color-octet $\Pi_a^0~[SU(2)_W$-
singlet] and $\Pi_a^\alpha~[SU(2)_W$-triplet]  composed of the techniquarks 
$~Q~$\footnote{These PGB's are the ones denoted by $\Theta_a^0$ and 
$\Theta_a^\alpha$ in Ref.\cite{AT}.} (the color-singlet PGB's in this
model are mainly composed of technileptons $L$, so that they are
irrelevant to the $t\bar{t}$ production). 
The decay constants of these  PGB's is $~F_Q =140~$GeV \cite{AT}. The
masses of these PGB's are model dependent \cite{AT}. In Ref.
\cite{AT}, the mass of $\Pi^\alpha_a$ is taken to be in the range
$~250~{\rm GeV}<m^\alpha_a<500~{\rm GeV}~$. We shall also take the mass
of $\Pi^0_a$ in the same range in our calculation.

 The relevant Feynman diagrams for the corrections to the $~\gamma\gamma
\rightarrow t\bar{t}~$ production amplitudes in the Appelquist-Terning model 
are shown in Fig.1(a)-(n). The Feynman rules needed in the calculations can 
be found in Ref.\cite{Feyn}. For instance, the  PGB-top (bottom) 
interactions are
\begin{eqnarray}                                    
\displaystyle  \frac{\sqrt{2}m_t}{f_Q}\big(
i\bar t\gamma_5 \frac{\lambda^a}{2}t\Pi^0_a+i\bar t\gamma_5 
\frac{\lambda^a}{2}t\Pi^3_a
+\frac{1}{\sqrt{2}}\bar t(1-\gamma_5) \frac{\lambda^a}{2}b\Pi^+_a
+\frac{1}{\sqrt{2}}\bar b(1+\gamma_5) \frac{\lambda^a}{2}t\Pi^-_a
\big).
\end{eqnarray}

In our calculation, we use dimensional regularization to 
regulate all the ultraviolet divergences in the virtual loop corrections and 
we adopt the on-mass-shell renormalization scheme. In that scheme, we 
need not consider the external top quark self energy diagrams. 
The renormalized amplitude 
for $~\gamma\gamma\rightarrow t \bar{t}~$ contains  
\begin{eqnarray}                                 
M_{ren} = M^{(t)}_{ren} + M^{(u)}_{ren}  
+\Delta M^{(\triangle)} \,,
\label{amp1}
\end{eqnarray} 
\noindent
where the superscripts $~t,u~$ stand for the $~t,u$-channel
amplitudes, respectively. $\Delta M^{(\triangle)}$ is the triangle 
correction of Fig.1 (n).  
Our notations for the momenta are:  
$~p_2~{\rm{and}}~p_1 $ denote the momenta of the outgoing $~t~$ and 
$~\bar{t}~$; $~p_3~$ and $~p_4~$ denote the momenta of the two incoming 
photons; $~\hat{s}\equiv (p_1+p_2)^2=(p_3+p_4)^2~$, $~\hat{t}\equiv 
(p_4- p_2)^2,~$ and $~\hat{u}\equiv (p_1 -p_4)^2~$. In eq.(1), 
$~M^{(t)}_{ren}~$ is contributed by Fig.1(a)-(m), and $~M^{(u)}_{ren}~$ is 
related to $~M^{(t)}_{ren}~$ by
\begin{eqnarray}                                 
M^{(u)}_{ren}& =& M^{(t)}_{ren}( p_3 \leftrightarrow p_4, \hat{t}
 \rightarrow \hat{u})\,.
\end{eqnarray}
The amplitude $~M^{(t)}_{ren}~$ contains
\begin{eqnarray}                                 
M^{(t)}_{ren} = M^{(t)}_0 +\Delta{ M^{(t)}}\,,     
\end{eqnarray} 
where
\begin{eqnarray}                                 
M^{(t)}_{0ij} &=& -i \frac{Q^2_te^2}{\hat{t}-m^2_t}\epsilon^{\mu}(p_4)
\epsilon^{\nu}(p_3)\bar{u}(p_2)\gamma_{\mu}(\gamma\cdot p_3 -\gamma
\cdot p_1+m_t) \gamma_{\nu}v(p_1)\delta_{ij}
\end{eqnarray}
is the tree-level $t$-channel amplitude, and $~\Delta{M^{(t)}}~$ is the
PGB correction to the $t$-channel amplitude which contains
\begin{eqnarray}                                 
\Delta{M^{(t)}}= \Delta M^{self(t)} +\Delta M^{v(t)}+ \Delta M^{b(t)}\,,
\end{eqnarray}
\noindent
in which $~\Delta M^{self(t)} $, $~\Delta M^{v(t)}~$, $~\Delta M^{b(t)}~$  
are the amplitudes contributed by the PGB's in the 
self-energy diagrams [Figs.1(b)-(c)], the vertex diagrams [Figs. 1(d)-(i)], 
the box diagrams [Figs.1(j)-(m)],
respectively.  They are

\begin{eqnarray}                                 
\Delta M^{self(t)}_{ij}& =& ie^2 \epsilon^{\mu}(p_4)\epsilon^{\nu}(p_3)
\bar{u}(p_2)[f^{self}_2\gamma_{\mu}\gamma_{\nu}+f^{self}_6 p_{2\mu} \gamma_{\nu} 
+f^{self}_{12}\rlap/p_4\gamma_{\mu}\gamma_{\nu}]v(p_1)\delta_{ij}\,,
\end{eqnarray}

\begin{eqnarray}                                  
\Delta M^{v(t)}_{ij} & = & ie^2 \epsilon^{\mu}(p_4)\epsilon^{\nu}(p_3)
\bar{u}(p_2)[f^v_2\gamma_{\mu}\gamma_{\nu}+f^v_3 p_{1\nu} \gamma_{\mu} 
+ f^v_6p_{2\mu}\gamma_{\nu}   \nonumber  \\
& &  + f^v_9 p_{2\mu}p_{1\nu}
 + f^v_{12}\rlap/p_4 \gamma_{\mu}\gamma_{\nu} + f^v_{13}\rlap/p_4 p_{1\nu}
\gamma_{\mu}+ f^v_{16}\rlap/p_4 p_{2\mu} \gamma_{\nu}]v(p_1)\delta_{ij}\,,
\end{eqnarray} 

\begin{eqnarray}                                    
\Delta M^{b(t)}_{ij}& =& ie^2 \epsilon^{\mu}(p_4)\epsilon^{\nu}(p_3)
\bar{u}(p_2)[f^b_1 g_{\mu \nu}+f^b_2 \gamma{\mu} \gamma_{\nu} 
+ f^b_3 p_{1\nu}\gamma_{\mu}   \nonumber \\
& & + f^b_4 p_{1\mu}\gamma_{\nu} + f^b_5 p_{2\nu}\gamma_{\mu}+ f^b_6 p_{2\mu}
 \gamma_{\nu} +f^b_7p_{1\mu}p_{1\nu} +f^b_8p_{1\mu}p_{2\nu}   \nonumber \\
& & + f^b_9 p_{2\mu}p_{1\nu} + f^b_{10}p_{2\mu}p_{2\nu}+ f^b_{11}\rlap/p_4
g_{\mu\nu}+ f^b_{12}\rlap/ p_4 \gamma_{\mu} \gamma_{\nu}    \nonumber  \\
& & + f^b_{13}\rlap/p_4 p_{1\nu}\gamma_{\mu} + f^b_{14} \rlap/p_4 p_{1\mu}
\gamma_{\nu}+ f^b_{15}\rlap/p_4 p_{2\nu}\gamma_{\mu}
+ f^b_{16}\rlap/p_4 p_{2\mu} \gamma_{\nu}   \nonumber \\
& & + f^b_{17}\rlap/ p_4 p_{1\mu} p_{1\nu} + f^b_{18} \rlap/p_4 p_{1\mu}
p_{2\nu}+ f^b_{19}\rlap/p_4 p_{2\mu}p_{1\nu}
+ f^b_{20}\rlap/p_4 p_{2\mu} p_{2\nu}]v(p_1)\delta_{ij}\,,
\end{eqnarray}
\noindent
and
\begin{eqnarray}                               
M^{(\triangle)}_{ren~ij}=ie^2\epsilon(p_4)\cdot\epsilon(p_3)
\bar{u}(p_2)f^\triangle v(p_1)
\delta_{ij}\,,
\end{eqnarray}
The explicit formulae for the form factors $~f$'s are 
given in the APPENDIX.

 The total cross section $~\sigma(s)~$ of the production of $~t\bar{t}~$ in 
$~\gamma\gamma~$ collisions is obtained by folding the the elementary cross 
section $~\sigma(\hat{s})~$ for the subprocess $~\gamma \gamma \to t \bar{t}~$ 
with the photon luminosity at the $ e^+ e^- $ colliders given in 
Ref.\cite{lumin}, i.e.
\begin{eqnarray}                                
\sigma(s)=\int^{x_{max}}_{2m_t/\sqrt{\hat{s}}}dz \frac{dL_{\gamma\gamma}} 
{d z}\sigma(\hat{s)}~~~~(\gamma\gamma\rightarrow t \bar{t}~~{\rm{at}}~~\hat{s}
=z^2 s) \,,
\end{eqnarray}
where $~\sqrt{s}(\sqrt{\hat{s}})~$ is the $~e^+e^- (\gamma \gamma)~$ center-of
-mass energy and $~dL_{\gamma \gamma}/dz~$ is the photon luminosity defined as
\begin{eqnarray}                                
\frac{dL_{\gamma\gamma}}{d z} = 2z \int^{x_{max}}_{z^2 /x_{max}}
 \frac{dx}{x}F_{\gamma/e}(x) F_{\gamma/e}(z^2/x)\,.
\end{eqnarray}
For unpolarized initial electrons and laser beams, the energy spectrum of the 
back-scattered photon is given by \cite{lumin} 
\begin{eqnarray}                                 
F_{\gamma/e}(x)  = \frac{1}{D(\xi)}[1 - x + \frac{1}{1-x} -\frac{4x}{
\xi(1-x)} + \frac{4x^2}{\xi^2(1-x)^2}]\,,
\end{eqnarray}
with
\begin{eqnarray}                                  
D(\xi)  =(1 - \frac{4}{\xi}-\frac{8}{\xi^2})\ln(1+\xi) + \frac{1}{2} 
+ \frac{8}{\xi} - \frac{1}{2(1+\xi)^2}\,,
\end{eqnarray}
where $~\xi = 4E_e\omega_0/ m_e^2$ in which $~m_e~$ and $~E_e~$ stand
,respectively, for the incident electron mass and energy, $~\omega_0~$ 
stands for the laser-photon energy, $~x=\omega /E_e $ stands for the fraction 
of energy of the incident electron carried by the back-scattered photon. 
$~F_{\gamma/e}(x)~$ vanishes for $~x>x_{max}=\omega_{max}/E_e
=\xi/(1+\xi)~$. In order to avoid the creation of $ e^+ e^- $ pairs by the 
interaction of the incident and backscattered photons, we require 
$~\omega_0 x_{max}\leq m_e^2/E_e~$ which implies $~\xi \leq 2+2\sqrt{2} 
\approx 4.8~$.  For the choice $~\xi = 4.8~$, we obtain 
\begin{eqnarray}                                   
x_{max} \approx 0.83, \hspace{2cm} D(\xi) \approx 1.8 \,.
\end{eqnarray}  

  In the calculation of $\sigma(\hat{s})$, instead of calculating
the square of the amplitude $~M_{ren}~$ analytically, we calculate the 
amplitudes numerically by using the method of Ref.\cite{HZ}. This greatly 
simplifies our calculations.  Care must be taken in the calculation of the 
form factors expressed in terms of the standard loop integrals defined in 
Ref.\cite{PV}. As has been discussed in Ref.\cite{Denner}, the formulae for 
the form factors given in terms of the tensor loop integrals will be ill-
defined when the scattering is forwards or backwards wherein the Gram
determinants of some matrices vanish and thus their inverse do not
exist. This problem can be solved by taking kinematic cuts on the
rapidity $~y~$ and the transverse momentum $~p_T~$. In this
paper, we take
\begin{eqnarray}                            
|y|< 2.5,~~~~~~~~~~p_T> 20~\rm{GeV}\,.
\end{eqnarray}
The cuts will also increase the relative correction \cite{Beenakker}.

 In our calculation, we take $m_t=176$~GeV, $m_b=4.9$~GeV, and we take
$\alpha_{em}(m_Z^2)=128.8$ with the one-loop running formula to determine the 
electromagnetic fine structure constant $\alpha_{em}$ at the desired scale.
The result of the tree-level cross sections are $\sigma_0=57.77$~fb for 
$\sqrt{s}=0.5$ TeV and $\sigma_0=535.4$~fb for $\sqrt{s}=1.5$ TeV. 
To see the main feature of the TC PGB corrections to the cross section, we
simply take $m_{\Pi^0_a}=m_{\Pi^3_a}=m_{\Pi^{\pm}_a}\equiv m_{\Pi_a}$ to
calculate the correction $\Delta \sigma$. The values of $\Delta\sigma$, 
the ratio $\Delta\sigma/\sigma_0$, and the total cross section 
$\sigma=\sigma_0+\Delta\sigma$ for $250~{\rm GeV}\leq
m_{\Pi_a}\leq 500~{\rm GeV}$ are listed in Table I. We see that for 
$\sqrt{s}=0.5$ TeV, the relative correction $\Delta\sigma/\sigma_0$ 
is of the order of ten percent which is about one order of magnitude 
larger than that in the MSSM with $\tan\beta\geq 1$
(which is about one percent) \cite{MSSM}. For 
$\sqrt{s}=1.5$~TeV, the relative corrections are around $(4-10)\%$ which is 
also larger than that in the MSSM with $\tan\beta\geq 1$.

For estimating the event rates, let us take an integrated luminosity of 
\begin{eqnarray}                             
\int {\cal L}dt=50~{\rm fb}^{-1},~~~~ {\rm for}~\sqrt{s}=0.5~{\rm TeV}\,,\nonumber\\
\int {\cal L}dt=100~{\rm fb}^{-1},~~~~ {\rm for}~\sqrt{s}=1.5~{\rm TeV}\,,
\end{eqnarray} 
which corresponds to a one year run at the NLC \cite{NLC}. There will be about 
$2500$ events for $\sqrt{s}=0.5$~TeV and $25000$ events for
$\sqrt{s}=1.5$~TeV according to the cross sections shown in 
Table I. The statistical uncertainty at $95\%$ C.L. is then
around $4\%$ for $\sqrt{s}=0.5$~TeV and $1.2\%$ for $\sqrt{s}=1.5$~TeV. 
Therefore {\it this model can be experimentally distinguished from the
MSSM model with $\tan\beta\geq 1$ in $\gamma\gamma\to t\bar{t}$ at the future 
photon collider}.\footnote{We only give here an order of magnitude
estimate considering only the statistical uncertainty. Practically,
the systematic error and the detection efficiency should also be 
taken into account which are beyond the scope of this paper.}

\vspace{0.5cm}
\begin{center}
 {\bf III. $t\bar{t}$ Production cross sections in TOPCTC Models }
\end{center}
\vspace{0.4cm}
\null\noindent
{\bf 1.}~The Original TOPCTC Model \cite{TOPC}

For TOPCTC models, we first consider the original TOPCTC model proposed
by Hill \cite{TOPC}. In this model, there are 60 TC PGB's in the TC sector
with the decay constant $f_{\Pi}=120~$GeV and three top-pions $\Pi_t^{0}$
, $\Pi_t^{\pm}$ in the topcolor sector with the decay constant
$f_{\Pi_t}=50~$GeV \cite{TOPC}. The top quark mass $m_t$ is mainly provided 
by the topcolor sector, while the TC sector only provide a small portion of it,
say $m'_t\sim 5-24$~GeV \cite{TOPC}\cite{Balaji}. The mass of the top-pion
depends on a parameter in the model \cite{TOPC}. For reasonable
values of the parameter, $m_{\Pi_t}$ is around $200~$GeV \cite{TOPC}.
The prediction of the light top-pion is the characteristic feature of
the TOPCTC models and this is the main difference between this kind of models 
and TC models without topcolor. In the following calculation, we 
would rather take a slightly larger range, $~180~{\rm GeV}\leq m_{\Pi_t}
\leq 300~$GeV, to see its effect, and we shall take the masses of the color-
singlet TC PGB's to vary in the range $100-325~$GeV.

The color-octet TC PGB-top (bottom) interactions are similar to eq.(1) but with
$m_t$ replaced by $m'_t$ and $f_Q$ replaced by $f_{\Pi}$, i.e.
\begin{eqnarray}                                    
\displaystyle  \frac{\sqrt{2}m'_t}{f_{\Pi}}\big(
i\bar t\gamma_5 \frac{\lambda^a}{2}t\Pi^0_a+i\bar t\gamma_5 
\frac{\lambda^a}{2}t\Pi^3_a
+\frac{1}{\sqrt{2}}\bar t(1-\gamma_5) \frac{\lambda^a}{2}b\Pi^+_a
+\frac{1}{\sqrt{2}}\bar b(1+\gamma_5) \frac{\lambda^a}{2}t\Pi^-_a
\big)\,.
\end{eqnarray}
The color-singlet TC PGB-top (bottom) interactions are \cite{Feyn}
\begin{eqnarray}                                    
\displaystyle  \frac{c_tm'_t}{\sqrt{2}f_{\Pi}}\big(
i\bar t\gamma_5 t\Pi^0+i\bar t\gamma_5 t\Pi^3
+\frac{1}{\sqrt{2}}\bar t(1-\gamma_5) b\Pi^+
+\frac{1}{\sqrt{2}}\bar b(1+\gamma_5) t\Pi^-
\big)\,,
\end{eqnarray}
where $\displaystyle c_t=\frac{1}{\sqrt{6}}$.
The interactions between the top-pions and the top (bottom) quark are 
\cite{Feyn}\cite{YZKL}
\begin{eqnarray}                                  
  \frac{m_t-m_t'}{\sqrt{2}f_{\Pi_t}}\big(
i\bar t\gamma_5 t\Pi_t^{0}
+\frac{1}{\sqrt{2}}\bar t(1-\gamma_5) b\Pi_t^{+}
+\frac{1}{\sqrt{2}}\bar b(1+\gamma_5) t\Pi_t^{-}
\big)\,.
\end{eqnarray}

The color-singlet TC PGB's and the top-pions can also couple to the two photons
via triangle fermion loops. It has been shown in Ref.\cite{ANOM} that, at the 
relevant energy, the technifermion triangle loops can be approximately 
evaluated from the formulae for the Adler-Bell-Jackiw anomaly \cite{ABJ} with 
which the general form of the effective $\Pi-B_1$-$B_2$ interaction ($\Pi$ 
denotes the PGB's $\Pi^0$, $\Pi^3$, $B_1$ and 
$B_2$ are gauge fields) is 
\cite{ANOM}
\begin{eqnarray}                                
\displaystyle
\frac{1}{(1+\delta_{B_1B_2})}\left(\frac{S_{\Pi B_1B_2}}{4
\sqrt{2}\pi^2F_\Pi}\right)\Pi
\epsilon_{\mu\nu\lambda\rho} ( \partial^\mu B^\nu_1)( \partial^\lambda 
B^\rho_2)\,,
\end{eqnarray}
where the factors $~S_{\Pi B_1B_2}~$ for various cases are given in Refs.
\cite{ANOM} and \cite{SYZ}. Note that the electromagnetic 
interaction violates $~SU(2)_W~$ symmetry, so that the
$\Pi^3-\gamma-\gamma$ and the $\Pi^0_t-\gamma-\gamma$ couplings do exist
 \cite{ANOM}. For example, from Ref.\cite{ANOM}, the $~S_{\Pi^0\gamma\gamma}~$ 
and $~S_{\Pi^3\gamma\gamma}~$ factors from the techniquark triangle
loop are 
\begin{eqnarray}                                 
S_{\Pi^0\gamma\gamma} & = & \frac{-4 e^2}{3\sqrt{6}}N_{TC} \,, \\
S_{\Pi^3\gamma\gamma} & = & \frac{4e^2}{\sqrt{6}}N_{TC} \,.
\end{eqnarray}
For the top quark triangle loop, the simple ABJ anomaly approach is not 
sufficient since the top quark mass is only $176~$GeV. Here we explicitly
calculate the top quark triangle loop and obtain the following
$\Pi^0_t-\gamma-\gamma$ interaction 
\begin{eqnarray}                                
-\frac{Q_t^2N_ce^2(m_t-m_t')}{16\pi^2\sqrt{2}f_{\Pi_t}}
(4m_tC_0)\Pi^0_t\epsilon_{\mu\nu\lambda\rho}(\partial^\mu A^\nu)
(\partial^\lambda A^\rho)\,, 
\end{eqnarray}
where $C_0(p_4,-p_4-p_3,m_t,m_t,m_t)$ is the standard 3-point Feynman integral
 \cite{PV}\footnote{From (19) we know that the $\Pi^0-t-\bar{t}$
coupling is proportional to $m'_t$, so that the top qurak triangle loop 
contribution to the $\Pi^0-\gamma-\gamma$ interaction is negligibly small.}.
 
From the above couplings, we see that there are additional important $s$-
channel diagrams contributing to the $t\bar{t}$ production shown in Fig.1(o)-
(p). The top-pion $s$-channel contribution is quite large compared with 
those from Fig.1(a)-(n) due to their strong coupling. This makes the 
contributions in this model quite different from those in the Appelquist-
Terning model presented in the last section.

Now we calculate the $s$-channel amplitude $~M^{(s)}_{ren}~$ in Fig.1(o)-(p). 
First of all, The $\Pi~(\Pi^0~{\rm{or}}~\Pi^3)$ propagator in Fig.1(o)-(p) 
takes the form
\begin{equation}                                
\frac{i}{\hat{s}-m_\Pi^2+im_\Pi\Gamma_\Pi} \,,
\end{equation}
where $\sqrt{\hat{s}}$ is the c.m. energy and $\Gamma_\Pi$ is the total width
of the PGB $\Pi$ which is important when $\sqrt{\hat{s}}$ is close to 
$m^2_\Pi$ (since $m_{\Pi_t}\sim 200~$GeV which is well below the
$t\bar{t}$ threshold, there is no need to include the width
$\Gamma_{\Pi_t}$ in the top-pion propagator). Similar to Ref. \cite{YZKL}, we 
can obtain the widths $\Gamma_{\Pi^0}$ and $\Gamma_{\Pi^3}$ which are 
\begin{equation}                                
\Gamma_{\Pi}=\Gamma(\Pi \rightarrow g_a g_b)+\Gamma(\Pi \rightarrow 
b\bar{b})
+\Gamma(\Pi \rightarrow t\bar{t})\,,\,\,\,\, {\rm{if}}~ m_{\Pi}>2m_t \,, 
\end{equation}
where
\begin{eqnarray}                                
\Gamma(\Pi^0\rightarrow g_a g_b) & = & \frac{\alpha _s^2c_t^2N_{TC}^2}{16 \pi ^2}
\frac{{m_{\Pi^0}}^3}{F_\Pi^2}\left|1 +\frac{J(R_{\Pi^0})}{2N_{TC}} \right|^2\,, \\
\Gamma(\Pi^3\rightarrow g_a g_b) & = & \frac{\alpha _s^2c_t^2}{64 \pi ^2}
\frac{{m_{\Pi^3}}^3}{F_\Pi^2}\left|J(R_{\Pi^3})\right|^2\,, \\
\Gamma(\Pi \rightarrow b\bar{b}) & = & \frac{3}{16 \pi}\frac{c_t^2m_b^2
m_{\Pi}}{F_\Pi^2}\sqrt{1-\frac{4m^2_b}{m^2_{\Pi}}} \,, \\
\Gamma(\Pi \rightarrow t\bar{t}) & = & \frac{3}{16 \pi}\frac{c_t^2m_t^2
m_{\Pi}}{F_\Pi^2}\sqrt{1-\frac{4m^2_t}{m^2_{\Pi}}} \,,\,\,\,\,\,{\rm{if}} ~
m_{\Pi}>2m_t\,, 
\end{eqnarray}
where \cite{SYZ}
\begin{eqnarray}                                 
J(R_\Pi)=-\frac{1}{R_\Pi^2}\int^1_0 \frac{dx}{x(1-x)}\ln[1-R_\Pi^2x(1-x)]\,,
\end{eqnarray}
in which $\displaystyle~R_\Pi\equiv \frac{m_\Pi}{m_t}~$.

With eqs.(18)-(31), we finally get
\begin{eqnarray}                                   
\Delta M^{(s)}_{ij} = ie^2 \epsilon^{\mu}(p_4)\epsilon^{\nu}(p_3)
\bar{u}(p_2)[f^s_{\mu\nu}\gamma_5]v(p_1)\delta_{ij}\,,
\end{eqnarray} 
where the form factor $~f^s_{\mu\nu}~$ is given in the APPENDIX.  

Let us first look at the contributions by different TC PGB's to the amplitude
$M^{(t,u,\triangle)}_{ren}$. From eq.(18) we see that, relative to the results 
in the last section, the color-octet TC PGB contributions to $M^{(t,u,
\triangle)}_{ren}$ from Fig.1(a)-(n) is suppressed by a factor $(\frac{m'_t/
f_\Pi}{m_t/f_Q})^2$ in this model. For $m'_t\sim 20~$GeV, this factor is about 
$2\%$ so that the color-octet TC PGB contributions are negligibly small in 
this model. From eqs.(19) and (18) we see that the color-singlet TC PGB 
contributions from Fig.1(a)-(n) is even smaller. For $m'_t=20~$GeV, the top-
pion contributions to $M^{(t,u)}_{ren}$ are not so small. The calculated 
results of $\Delta\sigma$ and $\Delta\sigma/\sigma_0$ from the contribution of 
$M^{(t,u)}_{ren}$ by the top-pions are listed in Table II. We see that 
these are slightly larger than those in Table I.

Numerical calculations show that the $\Pi^0$ and $\Pi^3$
contributions to the amplitude $M^{(s)}_{ren}$ from Fig.1(o)-(p) are also
negligiblly small, while the contribution from the top-pion $\Pi^0_t$ to 
$M^{(s)}_{ren}$ is quite large. Including this contribution, the
final results of $\Delta\sigma$ and the total cross section $\sigma=\sigma_0
+\Delta\sigma$ are listed in Table III. We see that these values of 
$\Delta\sigma$ are considerably larger than those listed in Table I.
Taking the integrated luminosity in eq.(17), we have around $1000$ events
for $\sqrt{s}=0.5$~TeV and around $40000$ events for $\sqrt{s}=1.5$~TeV.
The corresponding statistical uncertainties at the $95\%$ C.L. are then 
$6\%$ and $1\%$, respectively. Thus {\it this model is experimentally 
distinguishable from the Appelquist-Terning model and the MSSM with 
$\tan\beta\geq 1$ in $\gamma\gamma\to t\bar{t}$}.\\

\null\noindent
{\bf 2.}~The TOPCMTC Model \cite{Lane}\cite{YZKL}

This model is different from the original TOPCTC model \cite{TOPC}
mainly by the change of the TC sector in which $f_{\Pi}=40~$GeV instead of
the original $f_{\Pi}=120~$GeV, and \cite{Lane}
\begin{eqnarray}                                 
c_t&=&\frac{2}{\sqrt{6}}\,,\\
S_{\Pi^0\gamma\gamma} & = & \frac{10 e^2}{3\sqrt{6}}N_{TC} \,, \\
S_{\Pi^3\gamma\gamma} & = & \frac{2e^2}{\sqrt{6}}N_{TC} \,.
\end{eqnarray}
Thus, relative to the results in the last section, the suppression factor 
$(\frac{m'_t/f_{\Pi}}{m_t/f_Q})^2$ of the color-octet TC PGB contribution is 
now $0.16$ for $m'_t=20~$GeV, so that its contribution to $M^{(t,u,
\triangle)}_{ren}$ is still negligible (the relative correction is about 
$-0.2\%~{\rm to}~2\%$ for $\sqrt{s}=0.5$ TeV and $-0.1\%~{\rm to}~1\%$ for 
$\sqrt{s}=1.5$ TeV). The calculated results of the total $\Delta\sigma$ and 
$\sigma=\sigma_0+\Delta\sigma$ containing the contributions of top-pion 
from Fig.1(a)-(p) to both $M^{(t,u,\triangle)}_{ren}$ and $M^{(s)}_{ren}$ 
and of $\Pi^0,~\Pi^3$ from  Fig.1(o)-(p) to $M^{(s)}_{ren}$ for 
$m_\Pi\sim 100-325~$GeV are listed in Table IV. For $\sqrt{s}=0.5$ TeV,
we see that the values of $\Delta\sigma$ is slightly 
larger than those in Table III but not much. Thus the results of this
model are close to those in the original TOPCTC model at $\sqrt{s}=0.5$ TeV. 
For $\sqrt{s}=1.5$ TeV, the values of $\Delta\sigma$ in Table IV are
much larger than those in Table III, especially with large $m_t'$. 
To understand the reasons for such a difference, let us notice that the
main difference between these two topcolor assisted TC models comes from the
contributions of the PGB's $\Pi^0~{\rm and}~\Pi^3$ in the TC sector.
There are three factors in eq.(11) affecting this issue: 
 (a) in $\sigma(\hat{s})$,  $\Pi^0~{\rm and}~\Pi^3$
contributions are important at large  $\sqrt{\hat{s}}=m_{t\bar t}$
(cf. Fig.2); 
(b) the available range of $\sqrt{\hat{s}}$ is determined by
$min\{\sqrt{\hat{s}}\}=2m_t=352$~GeV and  $max\{\sqrt{\hat{s}}\}=x_{max}
\sqrt{s}=0.83\sqrt{s}$ (cf. eq.(15)), so that $max\{\sqrt{\hat{s}}\}$ is 
$s$-dependent;
(c) the $\gamma\gamma$ luminosity $\frac{dL_{\gamma\gamma}}
{dz}$ decreases rapidly in the vicinity of the $max\{\sqrt{\hat{s}}\}$ 
{cf. eq.(12))\cite{BZ} which gives rise to a suppression of the large 
$\sqrt{\hat{s}}$ contributions to $\sigma(s)$.
From (b) we see that the available $\sqrt{\hat{s}}$ in the case of 
$\sqrt{s}=0.5$~TeV is in a narrow range of 
$352-415$ GeV which is near the $t\bar{t}$ threshold, so that the 
contributions of $\Pi^0~{\rm and}~\Pi^3$ are less important than that of the 
top-pion (cf. (a)) and the additional $\gamma\gamma$ luminosity suppression
effect (c) plays a significant role in this narrow range. 
In the case of $\sqrt{s}=1.5$ TeV, the available $\sqrt{\hat{s}}$ spreads in 
a much wider range of $352-1245$ GeV which increases the importance of the 
$\Pi^0~{\rm and}~\Pi^3$ contributions (cf. (a)) and the $\gamma\gamma$ 
luminosity suppression effect (c) is less significant in this wide range.

Next we look at the total cross section $\sigma(s)$. Take the case of 
$m_{\Pi_t}=180$ GeV and $m_{\Pi}=100$ GeV as an example. The relative 
difference between the cross sections in the two tables for $m'_t=5$~GeV 
is about $6\%$, and for $m_t'=20$~GeV is about $15\%$ for $\sqrt{s}=0.5$~TeV 
and $17\%$ for $\sqrt{s}=1.5$~TeV. These are not so significant as the 
differences between the values of $\Delta\sigma$. To see the observability of 
the difference, we look at the statistical uncertainties.
Taking the integrated luminosity in eq.(17), we have $750-2000$ events 
for $\sqrt{s}=0.5$~TeV and $30000-40000$ events for $\sqrt{s}=1.5$~TeV 
according to the values of $\sigma$ given in Table IV. The statisical
uncertainties at the $95\%$ C.L. are thus $~(4-7)\%~$ for $\sqrt{s}=0.5$~TeV 
and around $1\%$ for $\sqrt{s}=1.5$~TeV. From this statistics, we first 
conclude that {\it this model can be clearly distinguished from the Appelquist-
Terning model and the MSSM with $\tan\beta\geq 1$ in the $\gamma\gamma\to 
t\bar{t}$ experiments}. 
Then we consider the relative differnce of $\sigma$ between these two topcolor 
assisted models, we see that the $6\%$ differnce in the case of $m'_t=5$~GeV 
can be easily observed at the $\sqrt{s}=1.5$~TeV collider, but is hard to be 
observed at the $\sqrt{s}=0.5$~TeV collider. In the case of $m'_t=20$~GeV, the 
$15\%$ and $17\%$ differences at $\sqrt{s}=0.5$~TeV and $1.5$~TeV can all be 
experimentally observed. Thus {\it even the difference between the original 
TOPCTC model and the TOPCMTC model can be clearly observed in the 
$\gamma\gamma\to t\bar{t}$ experiment at the $\sqrt{s}=1.5$~TeV photon 
collider}.

\null
\vspace{0.5cm}
\begin{center}
{\bf IV. Discussions and Conclusions}
\end{center}

In this paper, we have studied the possibility of testing different
currently interesting improved technicolor models in the $\gamma\gamma\to 
t\bar{t}$ experiments at the $\sqrt{s}=0.5$~TeV and $\sqrt{s}=1.5$~TeV photon 
colliders via the effects of their typical PGB's in the sense of the 
statistical uncertainty. In this calculation, we have neglected the corrections
from direct TC dynamics. Now we give an order of magnitude esatimate of the
TC dynamics effect based on the results in Ref.\cite{Asaka}. In Ref.
\cite{Asaka}, the authors considered the generation of the top quark
mass via ETC with a minimal TC model containing only one-doublet of 
technifermions. In such a simple model, there is no PGB. The ETC dynamics was
described by an effective four-fermion interaction between the technifermion 
and the top quark, and the coupling strength was determined by requiring that 
$m_t$ is completely given by this interaction. This four-fermion interaction 
provides a scalar TC-hadron (with mass around $2$ TeV) contribution to the 
$\gamma\gamma\to t\bar{t}$ cross section as a typical correction from the TC 
dynamics \cite{Asaka}. Such a TC-hadron correction only contributes in the 
case that the two photons are completely polarized in the same polarization 
\cite{Asaka}. So they considered only this completely polarized case in which 
the tree-level SM cross section (from $t$-channel top quark exchange) is 
suppressed at high energies \cite{DDS}. In addition, they imposed a polar 
angle cut $~\cos\theta_c=0.5~$ to further enhance the relative correction from 
the TC-hadron contribution. Their result is that the relative correction 
increases rapidly with the center-of-mass energy $\sqrt{\hat{s}}$: the 
relative correction at $\sqrt{\hat{s}}=0.5$~TeV is negligibly small, while 
those at $\sqrt{\hat{s}}=1~{\rm and}~1.5$~TeV are around $-8\%$ and $-43\%$, 
respectively \cite{Asaka}. What we have considered in this paper are the cross 
sections with unpolarized photons, and our polar angle cut is given in eq.(16) 
which corresponds to $~\cos\theta_c=0.99~$. We can estimate the corresponding 
TC-hadron relative correction in our case as follows. We know that averaging 
over the photon polarizations will reduce the TC-hadron contribution since it 
contributes only to a special polarization. On the other hand, the change of 
the polar angle cut from $~\cos\theta_c=0.5~$ to $~\cos\theta_c=0.99~$ will 
increase the TC-hadron contribution. As a rough order of magnitude estimate, 
we simply expect that ~$\left[\Delta\sigma^{had}_{TC}\right]^{unpolarized}_
{\cos\theta_c=0.99}\sim \left[\Delta\sigma^{had}_{TC}\right]^{polarized}_
{\cos\theta=0.5}$. 
Then
\begin{eqnarray}                                  
\displaystyle
\left[\frac{\Delta\sigma(\hat{s})^{had}_{TC}}{\sigma(\hat{s})_0}\right]^{
unpolarized}_{\cos\theta_c=0.99}\sim \left[\frac{\Delta\sigma(\hat{s})^{had}_
{TC}}{\sigma(\hat{s})_0}\right]^{polarized}_{\cos\theta_c=0.5}\cdot 
\frac{\left[\sigma(\hat{s})_0\right]^{polarized}_{\cos\theta_c=0.5}}{\left[
\sigma(\hat{s})_0\right]^{unpolarized}_{\cos\theta_c=0.99}}\,\,.
\end{eqnarray}
On the R.H.S. of eq.(36), the first factor is given in Ref.\cite{Asaka},
and the second factor can be calculated from the explicit formulae
given in Ref.\cite{DDS} which lead to $~\left[\sigma_0(\hat{s})\right]
^{pol}_{\cos\theta_c=0.5}/\left[\sigma_0(\hat{s})\right]^{unpolarized}
_{\cos\theta_c=0.99}=~0.42,~0.099,~0.039$ for $~\sqrt{\hat{s}}=~0.5,~1.0,
~{\rm and}~1.5~$TeV, respectively. From this we see that the present relative 
correction ~$\left[\Delta\sigma^{had}_{TC}(\hat{s})/\sigma(\hat{s})_0\right]
^{unpolarized}_{\cos\theta_c=0.99}$~ is also negliblely small at 
$\sqrt{\hat{s}}=0.5$~TeV, and those at $\sqrt{\hat{s}}=~1.0~{\rm and}~1.5$~TeV 
are $-0.8\%$ and $-1.7\%$, respectively. Taking the convolution (11) with the 
photon luminosity will not affect the order of magnitude. So that the
TC-hadron relative correction is of the same order of magnitude as the MSSM
corrections and is significantly smaller than the PGB corrections shown
in the last two sections. Thus we see that the correction from the direct TC
dynamics is at least not important in the present study\footnote{In the 
topcolor-assisted TC models the ETC only gives rise to a small portion $m'_t$ 
of the top quark mass ($m'_t\ll m_t$). So that the TC-hadron correction is 
even much smaller in such models.} and the conclusions in the last two 
sections are not affected by it. Our conclusions hold in most of the currently 
interesting improved TC model which contain certain PGB's. TC model without 
PGB does not belong to this category.

In summary, our calculation shows that \\
\null\noindent
(i)  corrections to the $t\bar{t}$ production cross sections in reasonable 
technicolor models without assited by topcolor are large enough to be 
observed, and are larger than those in the MSSM with $\tan\beta\geq 1$, 
so that these two kinds of models are experimentally distinguishable;\\
\null\noindent
(ii) corrections to the $t\bar{t}$ production cross sections in
topcolor assited technicolor models are much larger than those in
models without topcolor, and this kind of model can be clearly distinguished
from models without topcolor and MSSM with $\tan\beta\geq 1$ in the
$\gamma\gamma\to t\bar{t}$ experiments.\\
\null\noindent
(iii) it is even possible to distinguish the TOPCMTC model from the original
TOPCTC model in the $\gamma\gamma\to t\bar{t}$ experiment at the
$\sqrt{s}=1.5$~TeV photon collider, while these two models can be distinguished
at the $\sqrt{s}=0.5$~TeV photon collider only if the parameter $m'_t$
is as large as $20$~GeV.

We see that light neutral PGB contributions play an important role in the 
$\gamma\gamma\to t\bar{t}$ production rates. The direct detection of light 
neutral technipions at $e^+e^-$ colliders has been studied in Refs.
\cite{TC-pion} \cite{ANOM} and the signals are not so strong; the detection
at hadron colliders has been studied in Ref.\cite{Lane} and the
detection is more promising. The detection of the neutral top-pion has 
been discussed in Ref.\cite{TOPC} which shows that it imitates some effects 
of states in two-scale TC models. In Ref.\cite{YZKL}, it is shown that a 
neutral top-pion peak can be seen in the $t\bar{t}$ production at the LHC 
only if its mass is close to $350$~GeV. Our result in this paper provides 
an addtional test of the light neutral top-pion and technipion effects through 
the process $\gamma\gamma\to t\bar{t}$ at the LC.

We thus conclude that the $\gamma\gamma\to t\bar{t}$ experiments at the future 
photon colliders are really interesting in probing the electroweak symmetry
breaking mechanism.

\null
\noindent
{\bf ACKNOWLEDGMENT}

This work is supported by the National Natural Science Foundation of China, 
the Fundamental Research Foundation of Tsinghua University, a special
grant from the State Commission of Education of China, and the Natural 
Science Foundation of Henan Scientific Committee.

\null
\vspace{0.5cm}
\appendix
\section*{~~}

We present here the explicit formulae for the form factors appearing in
eqs.(7)-(10) and (32)\footnote{Here we have corrected some typographical mistakes
in the form factors given in a previous paper \cite{YKL}.}.
The renormalization constants are
\begin{eqnarray}
Z^k & = & B_1(p,m'_k,m_k)|_{p^2=m_t^2}
+2m^2_t \frac{\partial^{2}}{\partial p^{2}}(-Y_kB_0+B_1)
(p,m'_k,m_k)|_{p^2=m^2_t} \,, \\
\delta m^k & = & m_t[Y_kB_0-B_1](p,m'_k,m_k)|_{p^2=m^2_t} \,.
\end{eqnarray} 
\begin{eqnarray}
f^{self(b)(c)}_2 & = & \frac{Q_t^2}{(m^2_t-\hat{t})^2} 
\sum_{k=\Pi^0_t,\Pi^0_a,\Pi^3_a,\Pi^+_t, \Pi^+_a}q_k
\{2p_2.p_4[m_t(-Y_kB_0+Z^k)\\\nonumber
& & +\delta m^k+m_t(B_1-Z^k)]\} \,, \\
f^{self(b)(c)}_6 & = & \frac{Q_t^2}{(m^2_t-\hat{t})^2} 
\sum_{k=\Pi^0_t, \Pi^0_a,\Pi^3_a,\Pi^+_t, \Pi^+_a}q_k
\{-4m_t [m_t(-Y_kB_0+Z^k)+\delta m^k] \\ \nonumber
& & -4m_t^2 (B_1-Z^k)+4p_2.p_4(B_1-Z^k)\} \,, \\
f^{self(b)(c)}_{12} & = & \frac{1}{2} f^{self(b)(c)}_6 \,,
\end{eqnarray}
where $ B_0,B_1(t,m'_k,m_k) $ are 2-point Feynman integrals \cite{PV}, 
the superscripts $(b),~(c)$ indicate the labels in Fig.1 .

\begin{eqnarray}
f^{v(d)(e)}_2 & = & \frac{Q_t}{\hat{t}-m^2_t}\sum_{k=\Pi^0_t, \Pi^0_a,\Pi^3_a,\Pi_t^+,\Pi^+_a}Q'_kq_k
 [-2p_2.p_4 (m_t(C_0+C_{11})+m'_kY_kC_0)] , \\
f^{v(d)(e)}_6 & = & \frac{Q_t}{\hat{t}-m^2_t}\sum_{k=\Pi^0_t, \Pi^0_a,\Pi^3_a,\Pi_t^+,\Pi^+_a}Q'_kq_k
[1-4C_{24}- 2Z^k+m_t^2(2C_0+4C_{11}+2C_{21})\\ \nonumber
& & +2m'^2_kC_0+4m_tm'_kY_k(C_0+C_{11}) ] ,\\
f^{v(d)(e)}_{12} & = & \frac{Q_t}{\hat{t}-m^2_t}\sum_{k=\Pi^0_t, \Pi^0_a,\Pi^3_a,\Pi_t^+,\Pi^+_a}Q'_kq_k
[\frac{1}{2}-2C_{24}-Z^k+2p_2.p_4(C_{12}+C_{23})\\ \nonumber
& & m_t^2(C_0-C_{21})+m'^2_kC_0+2m_tm'_kY_kC_0] \,,\\
f^{v(d)(e)}_{16} & = & \frac{Q_t}{\hat{t}-m^2_t}\sum_{k=\Pi^0_t,\Pi^0_a,\Pi^3_a,\Pi_t^+,\Pi^+_a}Q'_kq_k
[-2m_t(C_{11}+C_{21})-2m_k'Y_kC_{11}]  \,,
\end{eqnarray}
where $C_0,C_{lm}(-p_2,p_4,m_k,m'_k,m'_k)$ are 3-point Feynman integrals.

\begin{eqnarray}
f^{v(f)}_6 & = & \frac{Q_t}{\hat{t}-m^2_t}\sum_{k=\Pi_t^+,\Pi^+_a}q_k
[-4C_{24}-2Z^k+4p_2.p_4(C_{12}+C_{23}) \\ \nonumber
& & -4m_t^2(C_{11}+C_{21}) ] ,\\
f^{v(f)}_{12} & = & \frac{Q_t}{\hat{t}-m^2_t}\sum_{k=\Pi_t^+,\Pi^+_a}q_k
[-2C_{24}-Z^k] ,\\
f^{v(f)}_{16} & = & \frac{Q_t}{\hat{t}-m^2_t}\sum_{k=\Pi_t^+,\Pi^+_a}q_k
[2m_t(C_{11}+C_{21})] \,,
\end{eqnarray}
where $C_0,C_{lm}(-p_2,p_4,m_b,m_k,m_k)$ are 3-point Feynman integrals.

\begin{eqnarray}
f^{v(g)(h)}_2 & = & \frac{Q_t}{\hat{t}-m^2_t}\sum_{k=\Pi^0_t, \Pi^0_a,\Pi^3_a,\Pi_t^+,\Pi^+_a}Q'_kq_k
 [-2m_tp_2.p_4 (C_0+C_{11})-2m_k'p_2.p_4Y_kC_0] , \\
f^{v(g)(h)}_3 & = & \frac{Q_t}{\hat{t}-m^2_t}\sum_{k=\Pi^0_t, \Pi^0_a,\Pi^3_a,\Pi_t^+,\Pi^+_a}Q'_kq_k
[4p_2.p_4(C_{12}+C_{23})] ,\\
f^{v(g)(h)}_6 & = & \frac{Q_t}{\hat{t}-m^2_t}\sum_{k=\Pi^0_t, \Pi^0_a,\Pi^3_a,\Pi_t^+,\Pi^+_a}Q'_kq_k
[1-4C_{24}-2Z^k+4p_2.p_4(C_{12}+C_{23}) \\ \nonumber
& & +m_t^2(2C_0-2C_{21})+2m_k'^2C_0+4m_tm_k'Y_kC_0]  ,\\
f^{v(g)(h)}_9 & = & \frac{Q_t}{\hat{t}-m^2_t}\sum_{k=\Pi^0_t, \Pi^0_a,\Pi^3_a,\Pi_t^+,\Pi^+_a}Q_k'q_k
[-4m_t(C_{11}+C_{21})] ,\\
f^{v(g)(h)}_{12} & = & \frac{1}{2}f^{v(g)(h)}_6\,,\\
f^{v(g)(h)}_{13} & = & \frac{1}{2}f^{v(g)(h)}_9\,,
\end{eqnarray}
where $C_0,C_{lm}(p_1,-p_3,m_k,m'_k,m'_k)$ are 3-point Feynman integrals.
\begin{eqnarray}
f^{v(i)}_3 & = & \frac{Q_t}{\hat{t}-m^2_t}\sum_{k=\Pi_t^+,\Pi^+_a}q_k
[-4p_2.p_4(C_{12}+C_{23})] ,\\
f^{v(i)}_6 & = & \frac{Q_t}{\hat{t}-m^2_t}\sum_{k=\Pi_t^+,\Pi^+_a}q_k
[-4C_{24}-2Z^k] , \\
f^{v(i)}_9 & = & \frac{Q_t}{\hat{t}-m^2_t}\sum_{k=\Pi_t^+,\Pi^+_a}q_k
[4m_t(C_{11}+C_{21})] ,\\
f^{v(i)}_{12} & = & \frac{1}{2}f^{v(i)}_{6},\\
f^{v(i)}_{13} & = & \frac{1}{2}f^{v(i)}_{9},
\end{eqnarray}
where $C_0,C_{lm}(p_1,-p_3,m_b,m_k,m_k)$ are 3-point Feynman integrals.

\begin{eqnarray}
f^{b(j)(k)}_1 & = & \sum_{k=\Pi^0_t, \Pi^0_a,\Pi^3_a,\Pi_t^+,\Pi^+_a}
Q'^2_kq_k[4m_t(D_{27} + D_{311})+4m_k'Y_kD_{27}],\\
f^{b(j)(k)}_2 & = &\sum_{k=\Pi^0_t, \Pi^0_a,\Pi^3_a,\Pi_t^+,\Pi^+_a}Q'^2_kq_k
[-m_t^3D_{31}+2m_t (p_2. p_4D_{34}+p_2. p_3D_{35}
     -p_3. p_4D_{310}\\ \nonumber
 & & -3D_{311})-6m_tD_{27}-m_t^3(3D_{21}+3D_{11}+D_0)
     +2m_tp_2. p_4(D_{21}+2D_{24} \\ \nonumber
 & & -D_{25}+D_{11}+D_{12}-D_{13})+2m_tp_2. p_3D_{25}-2m_tp_3. p_4 D_{26}\\ \nonumber
 & & +m_tm_k^{\prime 2}(D_0+D_{11})+m_k^\prime Y_k[(m^{\prime 2}_k -m_t^2)D_0
     -m^2_t(2D_{11}+D_{21})\\ \nonumber
 & & +2p_2. p_4(D_{11}+D_{12}-D_{13}+D_{24})+2p_2. p_3D_{25}
     -2p_3. p_4 D_{26}-4D_{27}]],\\
f^{b(j)(k)}_3 & = &   \sum_{k=\Pi^0_t, \Pi^0_a,\Pi^3_a,\Pi_t^+,\Pi^+_a}Q'^2_kq_k[2m_t^2 D_{12}-2m_k'^2 D_{12}-2m_t^2 D_{13}  \nonumber \\
& & +2m_k'^2 D_{13}+4m_t^2 D_{23}+4m_t^2 D_{24}-4m_t^2 D_{25}-4m_t^2 D_{26}+4D_{27}  \nonumber \\
& & -4m_t^2 D_{33}+2m_t^2 D_{34}-2m_t^2 D_{35}+4m_t^2 D_{37}+4m_t^2D_{39}-4m_t^2 D_{310}   \nonumber \\
& & +8D_{312}-12D_{313}+p_2 . p_4 (-4D_{22}-8D_{23}+4D_{25}+8D_{26}+4D_{33}-4D_{36} \nonumber \\
& & -4D_{37}+4D_{38}-8D_{39}+8D_{310})+p_1 . p_4 (-4D_{23}+4D_{26}+4D_{33}+4D_{38}-8D_{39})  \nonumber \\
& & +p_1 . p_2 (4D_{23}-4D_{26}-4D_{33}+4D_{37}+4D_{39}-4D_{310})], \\
f^{b(j)(k)}_4 & = &   \sum_{k=\Pi^0_t, \Pi^0_a,\Pi^3_a,\Pi_t^+,\Pi^+_a}Q'^2_kq_k[-2m_t^2 D_{13}-2m_k'^2 D_{13}+4m_t^2 D_{33}+2m_t^2 D_{35}  \nonumber \\
& & -4m_t^2 D_{37}+8D_{313}+p_2 . p_4 (4D_{23}-4D_{25}-4D_{33}+4D_{37}+4D_{39}-4D_{310})  \nonumber \\
& & +p_1 . p_4 (-4D_{33}+4D_{39})+p_1 . p_2 (4D_{33}-4D_{37})-4m_tm_k'Y_kD_{13}], \\
f^{b(j)(k)}_5 & = &   \sum_{k=\Pi^0_t, \Pi^0_a,\Pi^3_a,\Pi_t^+,\Pi^+_a}Q'^2_kq_k[-2m_t^2 D_{11}+2m_k'^2 D_{11}  \nonumber \\
& & +2m_t^2 D_{12}-2m_k'^2 D_{12}-4m_t^2 D_{21}+4m_t^2 D_{24}+4m_t^2 D_{25}  \nonumber \\
& & -4m_t^2 D_{26}-2m_t^2 D_{31}+2m_t^2 D_{34}+4m_t^2 D_{35}-4m_t^2 D_{37}  \nonumber \\
& & +4m_t^2 D_{39}-4m_t^2 D_{310}-8D_{311}+8D_{312}  \nonumber \\
& & +p_2 . p_4 (-4D_{22}+4D_{24}-4D_{25}+4D_{26}+4D_{34}-4D_{35}-4D_{36}  \nonumber \\
& & +4D_{37}+4D_{38}-4D_{39})+p_1 . p_4 (-4D_{25}+4D_{26}+4D_{37}+4D_{38}-4D_{39}-4D_{310}) \nonumber \\
& & +p_1 . p_2 (+4D_{25}-4D_{26}+4D_{35}-4D_{37}+4D_{39}-4D_{310})],\\
f^{b(j)(k)}_6 & = &   \sum_{k=\Pi^0_t, \Pi^0_a,\Pi^3_a,\Pi_t^+,\Pi^+_a}Q'^2_kq_k[
2m_t^2 D_0+2m_k'^2 D_0+4m_t^2 D_{11}  \nonumber \\
& & -4m_t^2 D_{13}+2m_t^2 D_{21}-4m_t^2 D_{25}+4D_{311}-4D_{313}  \nonumber \\
& & +p_1 . p_4 (4D_{25}-4D_{26})+4m_tm_k'Y_k(D_0+D_{11}-D_{13})],\\
f^{b(j)(k)}_7 & = &\sum_{k=\Pi^0_t, \Pi^0_a,\Pi^3_a,\Pi_t^+,\Pi^+_a}Q'^2_kq_k[
4m_t(D_{26}+D_{310})+4m_k'Y_kD_{26})],\\
f^{b(j)(k)}_8 & = & \sum_{k=\Pi^0_t, \Pi^0_a,\Pi^3_a,\Pi_t^+,\Pi^+_a}Q'^2_kq_k[
4m_t(-D_{25}+D_{26}-D_{35}+D_{310})+4m_k'Y_k(D_{26}-D_{25})],\\
f^{b(j)(k)}_9 & = & \sum_{k=\Pi^0_t, \Pi^0_a,\Pi^3_a,\Pi_t^+,\Pi^+_a}Q'^2_kq_k[
4m_t(-D_{12}-2D_{24}+D_{26}-D_{34}+D_{310})\\\nonumber
& & +4m_k'Y_k(D_{26}-D_{24}-D_{12})],\\
f^{b(j)(k)}_{10} & = & \sum_{k=\Pi^0_t, \Pi^0_a,\Pi^3_a,\Pi_t^+,\Pi^+_a}Q'^2_kq_k[
4m_t(D_{11}-D_{12}+2D_{21}-2D_{24}-D_{25}+D_{26}+D_{31} \nonumber  \\
& & -D_{34}-D_{35}+D_{310})+4m_k'Y_k(D_{11}-D_{12}+D_{21}-D_{24}
-D_{25}+D_{26})],\\
f^{b(j)(k)}_{11} & = &   \sum_{k=\Pi^0_t, \Pi^0_a,\Pi^3_a,\Pi_t^+,\Pi^+_a}Q'^2_kq_k[-4D_{27}-4D_{312}+4D_{313}],\\
f^{b(j)(k)}_{12} & = &   \sum_{k=\Pi^0_t, \Pi^0_a,\Pi^3_a,\Pi_t^+,\Pi^+_a}Q'^2_kq_k
[m_t^2(D_{34}-D_{35})+2[p_2. p_4(D_{310}-D_{36})
+p_2. p_3(D_{37}-D_{310})\\ \nonumber
 & & +p_3. p_4(D_{38}-D_{39})+4D_{312}-4D_{313}+2D_{27}]
     +m_t^2(2D_{24}-2D_{25}-D_{21})\\\nonumber
 & & +2[p_2. p_4(D_{25}+2D_{26}
    -D_{22}-D_{23})+p_2. p_3D_{25}-p_3. p_4 D_{26}]\\\nonumber
& &+m^2_t(D_0+D_{12}-D_{13})+m_k^{\prime 2}(D_0-D_{12}+D_{13})
+2m_tm_k^\prime Y_kD_0], \\
f^{b(j)(k)}_{13} & = & \sum_{k=\Pi^0_t, \Pi^0_a,\Pi^3_a,\Pi_t^+,\Pi^+_a}Q'^2_kq_k[
-2m_t(D_{11}+D_{24})-2m_k'Y_kD_{13}],\\
f^{b(j)(k)}_{14} & = & \sum_{k=\Pi^0_t, \Pi^0_a,\Pi^3_a,\Pi_t^+,\Pi^+_a}Q'^2_kq_k[
2m_t(D_{13}+D_{25})+2m_k'Y_kD_{13}],\\
f^{b(j)(k)}_{15} & = & \sum_{k=\Pi^0_t, \Pi^0_a,\Pi^3_a,\Pi_t^+,\Pi^+_a}Q'^2_kq_k[
2m_t(D_{11}-D_{12}+D_{21}-D_{24})+2m_k'Y_k(D_{11}-D_{12})],\\
f^{b(j)(k)}_{16} & = &\sum_{k=\Pi^0_t, \Pi^0_a,\Pi^3_a,\Pi_t^+,\Pi^+_a}Q'^2_kq_k
[2m_t(-D_{11}+D_{13}-D_{21}+D_{25})+2m_k'Y_k(D_{13}-D_{11})],\\
f^{b(j)(k)}_{17} & = & 4   \sum_{k=\Pi^0_t, \Pi^0_a,\Pi^3_a,\Pi_t^+,\Pi^+_a}Q'^2_kq_k[D_{23}-D_{26}-D_{38}+D_{39}],\\
f^{b(j)(k)}_{18} & = & 4   \sum_{k=\Pi^0_t, \Pi^0_a,\Pi^3_a,\Pi_t^+,\Pi^+_a}Q'^2_kq_k[D_{25}-D_{26}-D_{37}-D_{38}+D_{39}+D_{310}],\\
f^{b(j)(k)}_{19} & = & 4   \sum_{k=\Pi^0_t, \Pi^0_a,\Pi^3_a,\Pi_t^+,\Pi^+_a}Q'^2_kq_k[D_{22}+D_{23}-D_{25}-D_{26}+D_{36}-D_{38}+D_{39}-D_{310}],\\
f^{b(j)(k)}_{20} & = & 4   \sum_{k=\Pi^0_t, \Pi^0_a,\Pi^3_a,\Pi_t^+,\Pi^+_a}Q'^2_kq_k[D_{22}-D_{24}+D_{25}-D_{26}  \nonumber \\
& & -D_{34}+D_{35}+D_{36}-D_{37}-D_{38}+D_{39}] \,,
\end{eqnarray}
where $D_0,D_{lm},D_{lmn}(-p_2,p_4,p_3,m_k,m_k',m_k',m_k')$ are 4-point
Feynman integrals.
\begin{eqnarray}                                           
f^{b(l)}_1 & = & \sum_{k=\Pi_t^+,\Pi^+_a}q_k[-4m_t D_{311}],\\
f^{b(l)}_2 & = & 0,\\
f^{b(l)}_3 & = & 4 \sum_{k=\Pi_t^+,\Pi^+_a}q_k[D_{27}+D_{312}], \\
f^{b(l)}_4 & = & 4 \sum_{k=\Pi_t^+,\Pi^+_a}q_k[D_{313}],\\
f^{b(l)}_5 & = & 4 \sum_{k=\Pi_t^+,\Pi^+_a}q_k[-D_{311}+D_{312}],\\
f^{b(l)}_6 & = & 4 \sum_{k=\Pi_t^+,\Pi^+_a}q_k[-D_{27}-D_{311}+D_{313}],\\
f^{b(l)}_7 & = & -4m_t \sum_{k=\Pi_t^+,\Pi^+_a}q_k[D_{25}+D_{310}],\\
f^{b(l)}_8 & = & 4m_t \sum_{k=\Pi_t^+,\Pi^+_a}q_k[D_{35}-D_{310}],\\
f^{b(l)}_9 & = & 4m_t \sum_{k=\Pi_t^+,\Pi^+_a}q_k[D_{11}+D_{21}+D_{24}-D_{25}+D_{34}-D_{310}),\\
f^{b(l)}_{10} & = & 4m_t \sum_{k=\Pi_t^+,\Pi^+_a}q_k[-D_{21}+D_{24}-D_{31}
+D_{34}+D_{35}-D_{310}],\\
f^{b(l)}_{11} & = & \sum_{k=\Pi_t^+,\Pi^+_a}q_k[4D_{312}-4D_{313}],\\
f^{b(l)}_{12}& =& f^{b(l)}_{13}=f^{b(l)}_{14}=f^{b(l)}_{15}=f^{b(l)}_{16} = 0,\\
f^{b(l)}_{17} & = & 4 \sum_{k=\Pi_t^+,\Pi^+_a}q_k[-D_{23}+D_{26}+D_{38}-D_{39}],\\
f^{b(l)}_{18} & = & 4 \sum_{k=\Pi_t^+,\Pi^+_a}q_k[D_{37}+D_{38}-D_{39}-D_{310}],\\
f^{b(l)}_{19} & = & 4 \sum_{k=\Pi_t^+,\Pi^+_a}q_k[-D_{12}+D_{13}-D_{22}-D_{23}
-D_{24}+D_{25} \nonumber \\
& & +2D_{26}-D_{36}+D_{38}-D_{39}+D_{310}],\\
f^{b(l)}_{20} & = & 4 \sum_{k=\Pi_t^+,\Pi^+_a}q_k[-D_{22}+D_{24}-D_{25}+D_{26}  \nonumber \\
& & +D_{34}-D_{35}-D_{36}+D_{37}+D_{38}-D_{39}) \,, 
\end{eqnarray}
where $D_0,D_{lm},D_{lmn}(-p_2,p_4,p_3,m_b,m_k,m_k,m_k)$ are 4-point
Feynman integrals 
\begin{eqnarray}                                                   
f^{b(m)}_1 & = & 4m_t Q_b \sum_{k=\Pi_t^+,\Pi^+_a}q_k[-D_{27}-D_{311}+D_{313}],\\
f^{b(m)}_2 & = & 0,\\
f^{b(m)}_3 & = & Q_b \sum_{k=\Pi_t^+,\Pi^+_a}q_k[-2m_t^2D_{12}+2m_b^2D_{12}+2m_t^2D_{13}) \nonumber \\
& & -2m_b^2D_{13}-4m_t^2D_{24}+4m_t^2D_{25} \nonumber \\
& & -4D_{27}+2m_t^2D_{33}-2m_t^2D_{34}+2m_t^2D_{35} \nonumber \\
& & -2m_t^2D_{39}-8D_{312}+8D_{313} \nonumber \\
& & +p_2.p_4 (4D_{22}-4D_{26}+4D_{36}-4D_{310}) \nonumber \\
& & +p_1.p_2 (4D_{23}-4D_{26}+4D_{37}-4D_{310}) \nonumber \\
& & +p_1.p_4 (-4D_{23}+4D_{26}+4D_{38}-4D_{39})],\\
f^{b(m)}_4 & = & Q_b \sum_{k=\Pi_t^+,\Pi^+_a}q_k[-4D_{313}],\\
f^{b(m)}_5 & = & Q_b \sum_{k=\Pi_t^+,\Pi^+_a}q_k[2m_t^2D_{11}-2m_b^2D_{11}-2m_t^2D_{12} \nonumber \\
& & +2m_b^2D_{12}+4m_t^2D_{21}-4m_t^2D_{24} \nonumber \\
& & +2m_t^2D_{31}-2m_t^2D_{34}+2m_t^2D_{37}-2m_t^2D_{39}+8D_{311}
-8D_{312} \nonumber \\
& & +p_2.p_4 (4D_{22}-4D_{24}-4D_{34}+4D_{36}) \nonumber \\
& & +p_1.p_2 (4D_{25}-4D_{26}+4D_{35}-4D_{310}) \nonumber \\
& & +p_1.p_4 (-4D_{25}+4D_{26}+4D_{38}-4D_{310})], \\
f^{b(m)}_6 & = & -4 Q_b \sum_{k=\Pi_t^+,\Pi^+_a}q_k[D_{27}+D_{311}],\\
f^{b(m)}_7 & = & 4m_t Q_b \sum_{k=\Pi_t^+,\Pi^+_a}q_k[-D_{23}+D_{26}
+D_{33}-D_{37}-D_{39}+D_{310}],\\
f^{b(m)}_8 & = & 4m_t Q_b \sum_{k=\Pi_t^+,\Pi^+_a}q_k[-D_{25}+D_{26}
-D_{35}+D_{37}-D_{39}+D_{310}],\\
f^{b(m)}_9 & = & 4m_t Q_b \sum_{k=\Pi_t^+,\Pi^+_a}q_k[D_{12}-D_{13}+D_{23}+2D_{24} \nonumber \\
& & -2D_{25}-D_{26}+D_{34}-D_{35}+D_{37}-D_{310} ],\\
f^{b(m)}_{10} & = & 4m_t Q_b \sum_{k=\Pi_t^+,\Pi^+_a}q_k[-D_{11}+D_{12}-2D_{21}
+2D_{24} \nonumber \\
& & +D_{25}-D_{26}-D_{31}+D_{34}+D_{35}-D_{310}],\\
f^{b(m)}_{11} & = & 4 Q_b \sum_{k=\Pi_t^+,\Pi^+_a}q_k[D_{27}+D_{312}],\\
f^{b(m)}_{12} & = & Q_b \sum_{k=\Pi_t^+,\Pi^+_a}q_k[-2D_{27}],\\
f^{b(m)}_{13} & = & 2m_t Q_b \sum_{k=\Pi_t^+,\Pi^+_a}q_k[D_{12}-D_{13}+D_{23}+D_{24}
-D_{25}-D_{26}],\\
f^{b(m)}_{14} & = & 0,\\
f^{b(m)}_{15} & = & 2m_t Q_b \sum_{k=\Pi_t^+,\Pi^+_a}q_k[-D_{11}+D_{12}-D_{21}
+D_{24}+D_{25}-D_{26}],\\
f^{b(m)}_{16} & = & 0,\\
f^{b(m)}_{17} & = & 4 Q_b \sum_{k=\Pi_t^+,\Pi^+_a}q_k[D_{23}-D_{26}-D_{38}+D_{39}],\\
f^{b(m)}_{18} & = & 4 Q_b \sum_{k=\Pi_t^+,\Pi^+_a}q_k[D_{25}-D_{26}-D_{38}+D_{310}],\\
f^{b(m)}_{19} & = & 4 Q_b \sum_{k=\Pi_t^+,\Pi^+_a}q_k[-D_{22}+D_{26}-D_{36}+D_{310}],\\
f^{b(m)}_{20} & = & 4 Q_b \sum_{k=\Pi_t^+,\Pi^+_a}q_k[-D_{22}+D_{24}+D_{34}-D_{36}] \,,                                                                                                                                                     
\end{eqnarray}
where $D_0,D_{lm},D_{lmn}(-p_2,p_4,-p_1,m_k,m_b,m_b,m_k)$ are 4-point
Feynman integrals .
\begin{eqnarray}
& & f^\triangle = \sum_{k=\Pi_t^+,\Pi^+_a}q_k[2 m_t C_{11}
(-p_2,p_4+p_3,m_b,m_k,m_k)] .
\end{eqnarray}

\begin{eqnarray}
f^s_{\mu \nu}&=&\sum_{k=\Pi_t^0,\Pi^0,\Pi^3}
\frac{i\epsilon _{\mu \nu \rho \sigma}p_4^{\rho}p_3^{\sigma}}{\hat{s}-m^2_k+im_k\Gamma_k}
\big[8 N_c Q_t^2 q_k m_tC_0(p_4,-p_4-p_3,m_t,m_t,m_t)
-\frac{c_tm_t'{\cal S}_{k\gamma\gamma}}{8e^2\pi^2f_{\Pi}^2} \big] \,,
\end{eqnarray}

In the above: 
\begin{eqnarray}
& & Y_{\Pi^0_t}=Y_{\Pi^0_a}=Y_{\Pi^3_a}=-1,~~~Y_{\Pi^+_t}=Y_{\Pi^+_a}=0\\
\nonumber
& & m'_{\Pi^0_t}=m'_{\Pi^0_a}=m'_{\Pi^3_a}=m_t,
~~~m'_{\Pi^+_t}=m'_{\Pi^+_a}=m_b\\ \nonumber
& & Q'_{\Pi^0_t}=Q'_{\Pi^0_a}=Q'_{\Pi^3_a}=Q_t,
~~~Q'_{\Pi^+_t}=Q'_{\Pi^+_a}=Q_b\\ \nonumber
& & Q_t=\frac{2}{3}, \hspace{1cm} Q_b=-\frac{1}{3}, \hspace{1cm} N_c=3~.
\end{eqnarray}

In WTC model:
\begin{eqnarray}
& & q_{\Pi^{0}} = q_{\Pi^{3}}=q_{\Pi^{+}}=0~, \\
& & q_{\Pi^0_a}=q_{\Pi^3_a}=q_{\Pi^+_a} =\frac{2m_t^2}
{16\pi^2f_{\Pi}^2}\frac{4}{3}~,\\
& & q_{\Pi^0_t}=q_{\Pi_t^+}=0~,\\
& & {\cal S}_{\Pi^0\gamma\gamma}={\cal S}_{\Pi^3\gamma\gamma}=
{\cal S}_{\Pi_t^0\gamma\gamma}=0\;\;.
\end{eqnarray}

In TOPCTC model:
\begin{eqnarray}
& & q_{\Pi^{0}} = q_{\Pi^{3}}=q_{\Pi^{+}}=\frac{c_t^2m_t'^2}
{32\pi^2f_{\Pi}^2}~,  \\
& & q_{\Pi^0_a}=q_{\Pi^3_a}=q_{\Pi^+_a} =\frac{2m_t'^2}
{16\pi^2f_{\Pi}^2}\frac{4}{3}~,\\
& & q_{\Pi^0_t}=q_{\Pi_t^+}=\frac{(m_t-m_t')^2}
{32\pi^2f_{\Pi_t}^2}~,\\
& & c_t=\frac{1}{\sqrt{6}}~,
{\cal S}_{\Pi^0\gamma\gamma}=-\frac{4e^2N_{TC}}{3\sqrt{6}}\;\;,
{\cal S}_{\Pi^3\gamma\gamma}=\frac{4e^2N_{TC}}{\sqrt{6}}\;\;,
{\cal S}_{\Pi_t^0\gamma\gamma}=0\;\;.
\end{eqnarray}

In TOPCMTC model, the $q_k$s are the same as in (B) and 
\begin{eqnarray}
& & c_t=\frac{2}{\sqrt{6}}
\;\;,
{\cal S}_{\Pi^0\gamma\gamma}=\frac{10e^2N_{TC}}{3\sqrt{6}}\;\;,
{\cal S}_{\Pi^3\gamma\gamma}=\frac{2e^2N_{TC}}{\sqrt{6}}\;\;,
{\cal S}_{\Pi_t^0\gamma\gamma}=0\;\;.
\end{eqnarray}

\null

\newpage
{\bf Table I.}~~TC PGB corrections to the $\gamma\gamma\to t\bar{t}$ cross 
section $\Delta\sigma$, the relative correction $\Delta\sigma/\sigma_0$, and
the total cross section $\sigma=\sigma_0+\Delta\sigma$ for
various values of $m_{\Pi_a}$ in the Appelquist-Terning model
($\sigma_0=57.77$ fb for $\sqrt{s}=0.5$ TeV,
$\sigma_0=535.4$ fb for $\sqrt{s}=1.5$ TeV).\\

\begin{center}
\begin{tabular}{|c|c|c|c|c|c|c|}
\hline
 &\multicolumn{3}{|c|}{$\sqrt{s}=0.5~\rm{TeV}$} &
  \multicolumn{3}{|c|}{$\sqrt{s}=1.5~\rm{TeV}$}\\ \cline{2-7}
$m_{\Pi^a}$ (GeV) & $\Delta\sigma$ (fb) & $\Delta\sigma/\sigma_0$(\%) &
$~\sigma$ (fb) & $\Delta\sigma$ (fb) & $\Delta\sigma/\sigma_0$(\%) &
$~\sigma$ (fb)~~\\ \hline
250 & -9.11 & -15.8 & ~48.66 & -50.77~~& -9.5 & 484.6\\
275 & -8.22 & -14.2 & ~49.55 & -43.81~~& -8.2 & 491.6\\
300 & -7.47 & -12.9 & ~50.30 & -37.76~~& -7.1 & 497.6\\
325 & -6.83 & -11.8 & ~50.94 & -33.82~~& -6.3 & 501.6\\
350 & -6.28 & -10.9 & ~51.49 & -30.56~~& -5.7 & 504.8\\
375 & -5.80 & -10.0 & ~51.97 & -28.62~~& -5.3 & 506.8\\
400 & -5.38 & ~-9.3 & ~52.39 & -26.71~~& -5.0 & 508.7\\
425 & -5.02 & ~-8.7 & ~52.75 & -25.16~~& -4.7 & 510.2\\
450 & -4.69 & ~-8.1 & ~53.08 & -23.76~~& -4.4 & 511.6\\
475 & -4.40 & ~-7.6 & ~53.37 & -22.50~~& -4.2 & 512.9\\
500 & -4.13 & ~-7.1 & ~53.64 & -21.62~~& -4.0 & 513.8\\
\hline
\end{tabular}
\end{center}

\newpage
{\bf Table~II.}~~Top-pion contributions from Fig.1(a)-(n) to the 
$\gamma\gamma\to t\bar{t}$ production cross section $\Delta\sigma$, the
relative correction $\Delta\sigma/\sigma_0$ and
the total production cross section $\sigma=\sigma_0+\Delta\sigma$
in the original TOPCTC model ($\sqrt{s}=0.5$ TeV).

\begin{center}
\begin{tabular}{|c|c|c|c|}
\hline
$m_{\Pi_t}$ (GeV) & $\Delta\sigma$ (fb) & $\Delta\sigma/\sigma_0$(\%) 
& $\sigma$(fb) \\ \hline
180 & -13.04 & -22.6 & 44.73\\
185 & -9.81 & -17.0  & 47.96\\
190 & -8.05 & -13.9  & 49.72\\
195 & -7.27 & -12.6  & 50.05\\
200 & -7.22 & -12.5  & 50.55\\
210 & -7.86 & -13.6  & 49.91\\
215 & -7.83 & -13.6  & 49.94\\
225 & -7.57 & -13.1  & 50.20\\
250 & -6.73 & -11.6  & 51.04\\
275 & -5.97 & -10.3  & 51.80\\
300 & -5.35 & -9.3   & 52.42\\
\hline
\end{tabular}
\end{center}
\null
\vspace{1cm}

\newpage
{\bf Table~III.}~~Total TC PGB and top-pion contributions from Fig.1(a)-(p) to 
the $\gamma\gamma\to t\bar{t}$ production cross section $\Delta\sigma$ and
the total production cross section $\sigma=\sigma_0+\Delta\sigma$ in the
original TOPCTC model ($N_{TC}=4$, $\sigma_0=57.77$ fb for $\sqrt{s}=0.5$ TeV,
$\sigma_0=535.4$ fb for $\sqrt{s}=1.5$ TeV,).\\

\begin{center}
\begin{tabular}{|c|c|c|c|c|c|c|c|c|}
\hline
 & \multicolumn{4}{|c|}{$\sqrt{s}=0.5~\rm{TeV}$} &
 \multicolumn{4}{|c|}{$\sqrt{s}=1.5~\rm{TeV}$}\\ \cline{2-9}
 $m_{\Pi_t}$ (GeV) & 
 \multicolumn{2}{|c|}{$m_t'=5$ GeV} &
 \multicolumn{2}{|c|}{$m_t'=20$ GeV} & 
 \multicolumn{2}{|c|}{$m_t'=5$ GeV} &
 \multicolumn{2}{|c|}{$m_t'=20$ GeV} \\
 \cline{2-9}
& $\Delta\sigma$ (fb) & $\sigma$ (fb) & $\Delta\sigma$ (fb) & $\sigma$ (fb)
& $\Delta\sigma$ (fb) & $\sigma$ (fb) & $\Delta\sigma$ (fb) & $\sigma$ (fb)\\ 
\hline
180 & -41.96 & 15.81 & -38.53 & 19.24& -179.5 & 355.9 & -154.7 & 380.7\\
200 & -37.70 & 20.07 & -35.13 & 22.64& -145.3 & 390.1 & -125.3 & 410.1\\
225 & -40.53 & 17.24 & -38.04 & 19.73& -123.9 & 411.5 & -108.2 & 427.2\\
250 & -41.27 & 16.50 & -39.68 & 18.09& -113.8 & 421.6 & -99.3 & 436.1\\
275 & -40.34 & 17.43 & -40.87 & 16.90& -107.0 & 428.4 & -96.1 & 439.3\\
300 & -33.37 & 22.40 & -38.97 & 18.80& -101.9 & 433.5 & -94.2 & 441.2\\
\hline
\end{tabular}
\end{center}

\newpage
{\bf Table~IV.}~Total TC PGB and top-pion contributions from Fig.1(a)-(p) to 
the $\gamma\gamma\to t\bar{t}$ production cross section $\Delta\sigma$ and
the total production cross section $\sigma=\sigma_0+\Delta\sigma$ in the
TOPCMTC model ($N_{TC}=4$, $\sigma_0=57.77$ fb for $\sqrt{s}=0.5$ TeV,
$\sigma_0=535.4$ fb for $\sqrt{s}=1.5$ TeV).\\

\begin{center}
\begin{tabular}{|c|c|c|c|c|c|c|c|c|}
\hline
\multicolumn{9}{|c|}{$\sqrt{s}=0.5~\rm{TeV}$}\\ \hline
$m_{\Pi^0,\Pi^3}$ & \multicolumn{4}{|c|}{$m_t'=5$ GeV}
& \multicolumn{4}{|c|}{$m_t'=20$ GeV} \\ \cline{2-9}
& \multicolumn{2}{|c|}{$m_{\Pi_t}=180$ GeV} 
& \multicolumn{2}{|c|}{$m_{\Pi_t}=250$ GeV} 
& \multicolumn{2}{|c|}{$m_{\Pi_t}=180$ GeV} 
& \multicolumn{2}{|c|}{$m_{\Pi_t}=250$ GeV}\\ \cline{2-9}
 & $\Delta\sigma$ (fb) & $\sigma$ (fb) & $\Delta\sigma$ (fb) & $\sigma$ (fb)
 & $\Delta\sigma$ (fb) & $\sigma$ (fb) & $\Delta\sigma$ (fb) & $\sigma$ (fb)\\ 
 \hline
100 & -42.88 & 14.89 & -41.35 & 16.42 & -42.92 & 14.85 & -41.30 & 16.47\\
150 & -42.95 & 14.82 & -41.33 & 16.44 & -43.03 & 14.74 & -41.12 & 16.65\\
200 & -43.06 & 14.71 & -41.29 & 16.48 & -43.06 & 14.71 & -40.63 & 17.14\\
250 & -43.23 & 14.54 & -41.17 & 16.60 & -42.54 & 15.23 & -39.07 & 18.70\\
300 & -43.43 & 14.34 & -40.51 & 17.26 & -38.11 & 19.66 & -32.19 & 25.58\\
325 & -43.24 & 14.53 & -39.30 & 18.47 & -26.16 & 31.61 & -17.22 & 40.55\\
\hline
\multicolumn{9}{|c|}{$\sqrt{s}=1.5~\rm{TeV}$}\\ \hline
$m_{\Pi^0,\Pi^3}$ & \multicolumn{4}{|c|}{$m_t'=5$ GeV}
& \multicolumn{4}{|c|}{$m_t'=20$ GeV} \\ \cline{2-9}
& \multicolumn{2}{|c|}{$m_{\Pi_t}=180$ GeV} 
& \multicolumn{2}{|c|}{$m_{\Pi_t}=250$ GeV} 
& \multicolumn{2}{|c|}{$m_{\Pi_t}=180$ GeV} 
& \multicolumn{2}{|c|}{$m_{\Pi_t}=250$ GeV}\\ \cline{2-9}
 & $\Delta\sigma$ (fb) & $\sigma$ (fb) & $\Delta\sigma$ (fb) & $\sigma$ (fb)
 & $\Delta\sigma$ (fb) & $\sigma$ (fb) & $\Delta\sigma$ (fb) & $\sigma$ (fb)\\ 
 \hline
100 & -201.2 & 334.2 & -136.9 & 398.5 & -218.3 & 317.1 & -166.7 & 368.7\\
150 & -201.8 & 333.6 & -137.7 & 397.7 & -218.3 & 317.1 & -166.1 & 369.3\\
200 & -203.4 & 332.0 & -138.2 & 397.2 & -219.3 & 316.1 & -166.9 & 368.5\\
250 & -204.3 & 331.1 & -139.0 & 396.4 & -220.0 & 315.4 & -167.2 & 368.2\\
300 & -206.7 & 328.7 & -140.8 & 394.6 & -219.9 & 315.5 & -166.0 & 369.4\\
325 & -207.1 & 328.3 & -141.3 & 394.1 & -212.7 & 322.7 & -157.8 & 377.6\\
\hline
\end{tabular}
\end{center}

\newpage
\begin{center}
\epsfysize=20cm
\epsfig{file=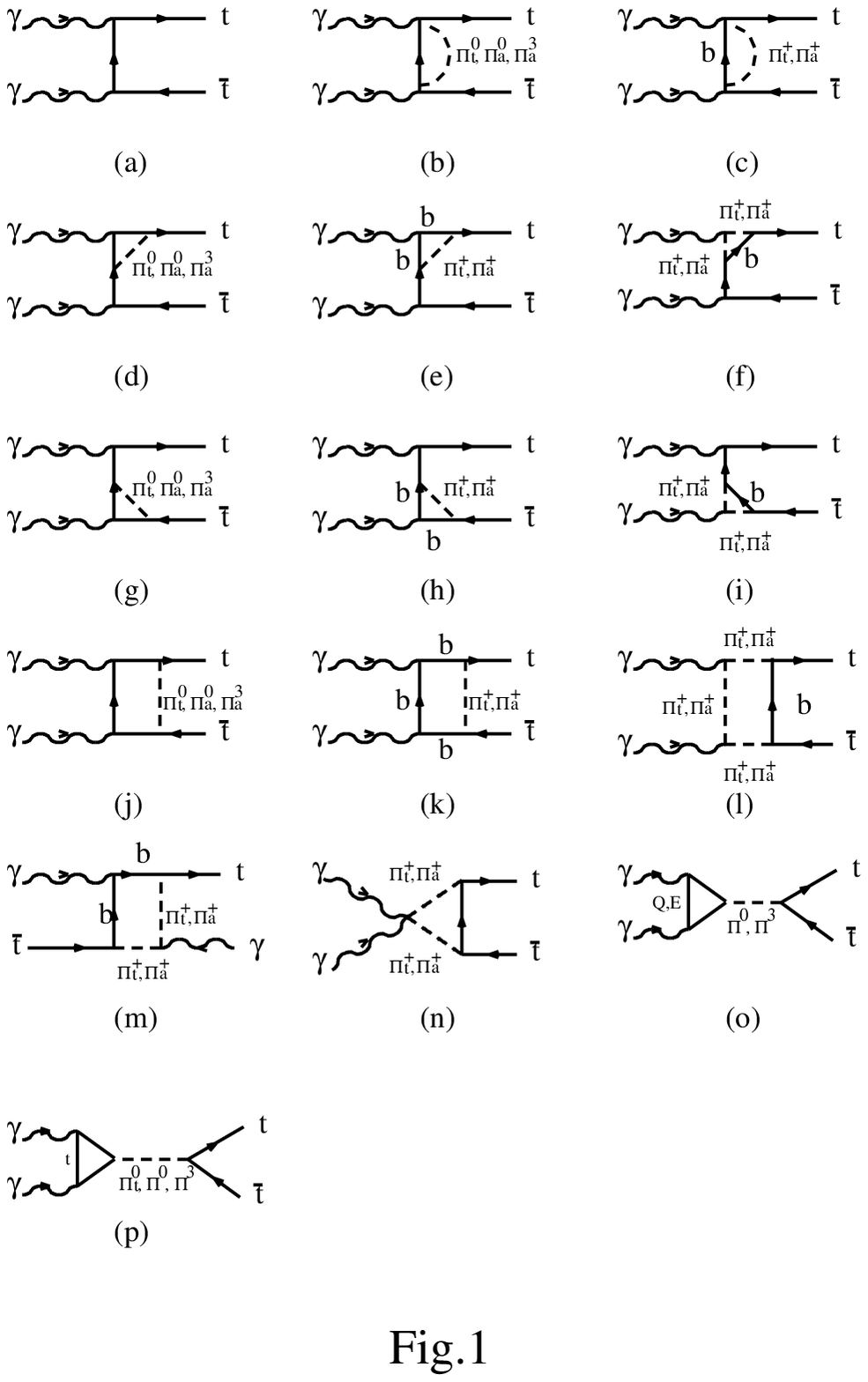}
{\bf Fig. 1 } Feynman diagrams for PGB contributions to the $~\gamma 
\gamma \to t \bar{t}~$ process. 
(a): tree level diagrams; (b)-(c): self-energy diagams; (d)-(i): vertex 
diagams;  (j)-(m): box  diagams; (n): triangle diagram; (o)-(p):
$s$-channel diagrams.
\end{center}

\newpage
\begin{center}
\epsfysize=20cm
\epsfig{file=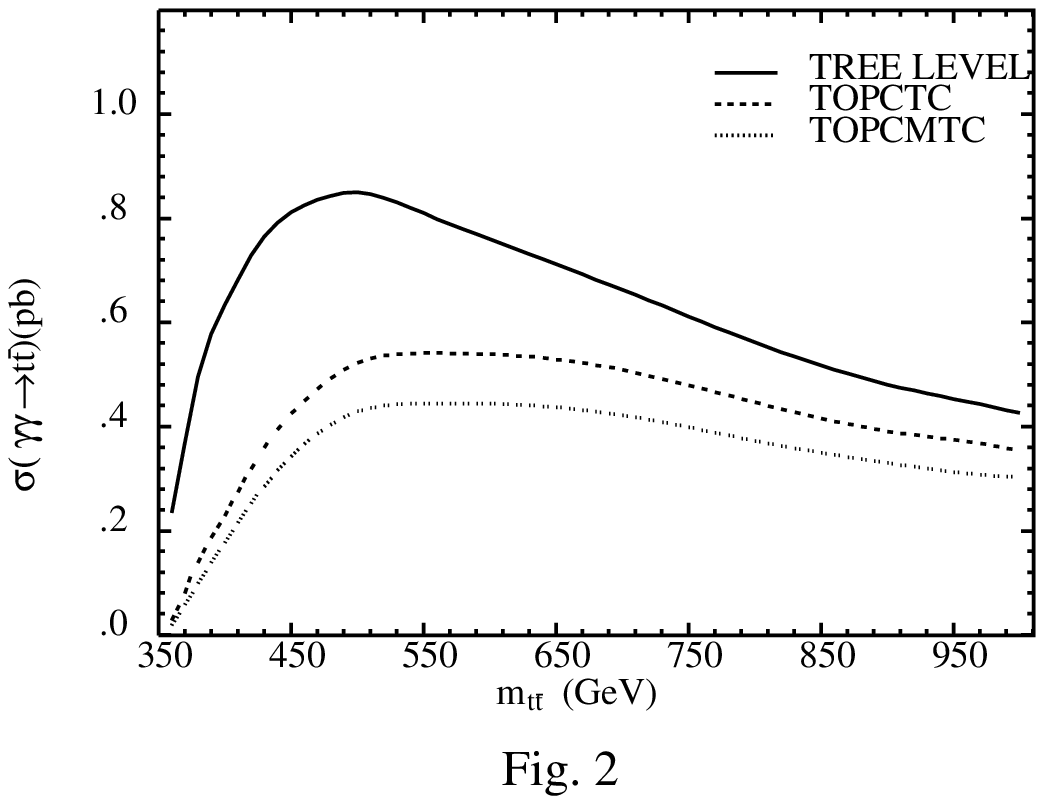}
{\bf Fig. 2 } Subprocess cross sections $\sigma(\hat{s})$ in the tree
level SM, the original TOPCTC model and the TOPCMTC model with  
$m_{\Pi_t}=180$ GeV, $m_{\Pi^0,\Pi^3}=100$ GeV and $m'_t=20$ GeV.
\end{center}

\end{document}